\documentclass[amsmath,twocolumn,showpacs,aps,prl]{revtex4}
\usepackage{bm}
\usepackage{graphics}
\usepackage{graphicx}

\begin{document}

\title{Non-equilibrium Spin Waves in Paramagnetic Metals}

\author{A. A. Zyuzin and A. Yu. Zyuzin}

\affiliation{A.F. Ioffe Physico-Technical Institute of Russian
Academy of Sciences, 194021 St. Petersburg, Russia}


\begin{abstract} We theoretically study the effect of exchange
interaction on the non-equilibrium spin waves in disordered
paramagnetic metals under the spin injection condition. We show
that the gapless spectrum of spin waves, describing the spin
precession in the absence of the applied magnetic field, changes
sign to negative on the paramagnetic side near the ferromagnet -
paramagnet phase transition. The damping of spin waves is small in
the limit when electron-electron exchange energy is larger than
the inverse electron mean free time, while in the opposite limit
the propagation of spin waves is strongly suppressed. We discuss
the amplification of the electromagnetic field by the
non-equilibrium spin waves.
\end{abstract}

\maketitle

\section{Introduction} Effects related to non-equilibrium spin
accumulation in metals and semiconductors attract considerable
interest in the field of spin phenomena (see for a review
\cite{bib:Dyakonov1, bib:Review}). The spin accumulation in
nonmagnetic materials can be produced by different types of
pumping such as, for example, optical orientation and electrical
methods. In the optical spin orientation method \cite{bib:Optical}
the electron's orbital momentum is oriented by the absorption of
circularly polarized electromagnetic field and as a result spin
becomes polarized through the spin-orbit interactions. The
electrical spin injection method is based on transfer of
spin-polarized electrons from the ferromagnet into the paramagnet.
The mechanism of spin injection from ferromagnet into
semiconductors and metals was first proposed by Aronov and Pikus
\cite{bib:Aronov1, bib:Aronov2}. It was shown that the electrical
current through the contact between a ferromagnet and a paramagnet
will produce the non-equilibrium magnetization in the paramagnetic
region on the scale of the spin diffusion length. The
non-equilibrium spin magnetization was experimentally studied in
metals by Johnson and Silsbee \cite{bib:JS1, bib:JS2} by voltage
measurement.

Despite of extensive study of various aspects of spin dynamics in
paramagnets \cite{bib:Dyakonov1, bib:Review}, little attention was
paid to the effect of electron-electron exchange interaction on
the non-equilibrium spin transport. Aronov \cite{bib:Aronov}
predicted the existence of the long-lived spin waves excitations
in the clean paramagnets with the non-equilibrium spin
polarization under spin injection. The spectrum of these spin
waves has a quadratic power low dependence on the wave vector
$\propto q^2$ at small $q$, while the Landau damping of spin waves
is small compared to theirs frequency. Interestingly, spin waves
in non - equilibrium system have negative frequency, which means
that the excitation of these spin waves lowers the energy of the
system. Motivated by this research, we extend the study to the
case of strongly enhanced and disordered paramagnetic metals with
the non-equilibrium spin polarization.

Berger \cite{bib:Berger} and Slonczewski \cite{bib:Slonczewski}
calculated the spin-transfer torque effect in thin film
ferromagnetic-nonmagnetic metal multilayer system. They predicted
the stimulated emission of spin waves in ferromagnetic layers due
to \emph{negative} Gilbert damping at high values of current
density applied to the multilayer.  M. Tsoi et al. \cite{bib:Tsoi}
first observed such current-induced excitations in a magnetic
multilayer in magnetoresistance measurements. Recently, the direct
measurements of both the magnitude and the direction of the
spin-transfer torque in a magnetic tunnel junction were given in
papers (see \cite{bib:Sankey, bib:Kubota} and references therein).
We argue that the effects related to the non-equilibrium spin
excitations in the paramagnetic layers also might become
interesting.

In the present paper, we show that in contrast to the stimulated
emission of spin waves in the ferromagnetic layers of the magnetic
multilayer systems \cite{bib:Berger, bib:Slonczewski}, the
non-equilibrium transverse spin waves in paramagnetic metals are
stable under spin injection. On the other hand the excitation of
these spin waves decreases the energy of the system. Indeed, we
found that there is a crossover from positive to negative sign of
the frequency in the vicinity of the ferromagnet - paramagnet
phase transition. The real part of the transverse spin waves
frequency in the paramagnetic metal has negative sign, while the
imaginary part describing the damping of spin waves is negative.
We found that in the regime when the exchange energy of electrons
is smaller than the inverse mean free time the propagation of spin
waves is strongly suppressed. Interestingly, the electromagnetic
field might be amplified via coupling to the non-equilibrium spin
waves. Clearly, one deals with the system driven by the external
spin injection source to the highly excited spin correlated state.
The energy might be extracted from this state via creation of the
spin waves. Based on these results, we propose an electromagnetic
field amplification in the spin-wave resonance experiment.

\section{Definitions} Let us consider the tunneling contact
between a ferromagnetic metal with spontaneous magnetization and a
paramagnetic metal. The current flow through the contact produces
the non-equilibrium magnetization \cite{bib:Aronov2} in the
paramagnetic region. The realization of spin injection is well
studied experimentally \cite{bib:JS1}. In addition, we take into
account the external magnetic field applied to the paramagnet
leading to the Zeeman effect. For simplicity, we assume the
direction of magnetic field is parallel to the non-equilibrium
magnetization in the paramagnetic metal. We suggest the size of
the paramagnet is much smaller than the typical spin relaxation
length corresponding, for example, to the spin-orbit scattering.
Therefore, the electron interactions with impurities are assumed
to be spin independent.

\begin{figure}[t] \centering
\includegraphics[width=8cm]{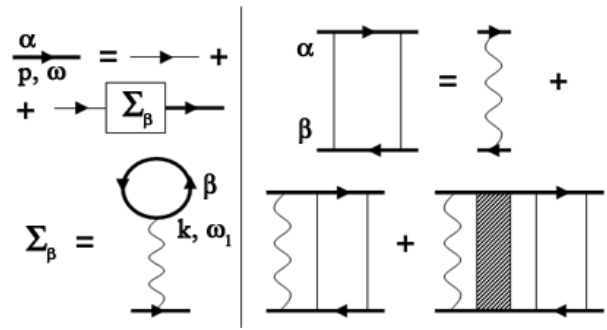}
\caption{Left: Diagram equation for a Green's function and self
energy exchange diagram. Right: ladder approximation for the
magnetic susceptibility, where shaded area stands for the diffuson
ladder, see \cite{bib:AA}. $\alpha$ and $\beta$ denote spin up or
spin down.}\label{fig:1}
\end{figure}

In order to obtain the spectrum of transverse spin waves in the
paramagnet we will calculate the poles of the magnetic
susceptibility using Keldysh formalism. The equation for the
Keldysh function $G_K$ and the retarded and advanced Green's
functions $G_R$, $G_A$ is presented in Fig. (\ref{fig:1}, left).

The retarded and advanced Green's functions in the Fourier
representation are given by
\begin{equation}\label{RA}
G^{\uparrow ,\downarrow}_{R,A}(p,\omega) =[\omega - E^{\uparrow
,\downarrow}_p +\mu \pm i/2\tau ]^{-1}
\end{equation}
where $\tau$ is the mean free time and $\mu$ is the chemical
potential. The energy of spin up and spin down electrons is given
as
\begin{equation}\label{energy}
E^{\uparrow ,\downarrow}(\mathbf{p})=\frac{p^2}{2m}
\mp\omega_z - \frac{i}{2}\lambda
\int\frac{d\mathbf{k}d\omega}{(2\pi)^4}G_K^{ \downarrow ,
\uparrow}(\mathbf{k},\omega)
\end{equation}
where $\lambda$ is the electron-electron exchange coupling
constant. Last term in expression (\ref{energy}) describes
contribution of the short-range electron-electron exchange
interactions to the spin splitting, Fig. (\ref{fig:1}, Left), see
\cite{bib:White}. "Hartree term" has negative sign and consists of
two contributions corresponding to two spin up and down electron
states at the bubble. "Fock term" to the self energy of electrons
is positive and has one spin contribution which cancels Hartree
contribution with equal spins. Correspondingly, there is only one
term left, shown in Fig. (\ref{fig:1}, Left), with different spins
denoted as $\alpha$ and $\beta$. We assume that the free electron
dispersion is parabolic $p^2/2m$ with an effective mass $m$. We
also introduce here an external magnetic field which contributes
the Zeeman energy $\omega_z = \mu_B H_z$ to definition
(\ref{energy}). We suggest that the magnetic field is applied in
the $\mathbf{z}$-direction and parallel to the non-equilibrium
magnetization in the paramagnet.

Spin-polarized current flow through the ferromagnet - paramagnet
interface leads to the different densities of spin up and spin
down electrons in the paramagnetic metal \cite{bib:Aronov2} on the
scale of spin relaxation length. Since we assume the spin
relaxation processes to be weak, we suggest the system under
consideration is in the energy equilibrium state while having
non-equilibrium spin accumulation. The resulting spin accumulation
is uniform over the paramagnetic sample and is proportional to the
value of spin-polarized current. The electron chemical potential
shifts for spin up and spin down electrons
$\mu^{\uparrow,\downarrow} = \pm \delta \mu /2$. Then the Keldysh
function does not depend on the spacial coordinates of the system
and one has
\begin{equation}\label{GK}
G^{\uparrow ,\downarrow}_K(p,\omega) = (1-2f^{\uparrow
,\downarrow}(\omega))[G^{\uparrow
,\downarrow}_R(p,\omega)-G^{\uparrow ,\downarrow}_A(p,\omega)]
\end{equation}
where Fermi distribution function at a given temperature $T$ is
$f^{\uparrow ,\downarrow}(\omega) = (e^{(\omega - \mu^{\uparrow
,\downarrow})/T} + 1)^{-1}$. Due to the electron exchange
interaction the energy of electrons depends on the non-equilibrium
chemical potential shifts.

Ladder approximation \cite{bib:White, bib:Moriya, bib:AA} for the
transverse dynamical magnetic susceptibility is shown in Fig.
(\ref{fig:1}, right) where we assume short range electron-electron
interactions.
The spectrum of non-equilibrium spin waves is determined by the
poles of the susceptibility $\chi^{+-}(q,\Omega)$ for the circular
spin component $s^{+} = s_{x} + i s_{y}$.

The dynamical magnetic susceptibility can be evaluated in the
leading order in $1/\mu\tau$ by summing the ladder vertex
correction  to the polarization bubble due to exchange
interactions between electrons and impurity scattering. The
general result for the dynamical susceptibility takes the form
\begin{equation}\label{suscept}
\chi^{+-}(q,\Omega) = 2\mu^2_B \nu
\frac{\Pi(q,\Omega)}{1-\lambda\nu\Pi(q,\Omega)}
\end{equation}
where $\nu$ is the electron density of states per spin direction
and $\Pi(q,\Omega)$ is determined through the non-equilibrium
Green's functions as
\begin{eqnarray}\nonumber
\Pi(q,\Omega)= \frac{i}{2\nu}\int\frac{d\mathbf{p} d\omega
}{(2\pi)^4} [\frac{G_{R}^{ \downarrow}(p_+,\omega_+)G_{K}^{
\uparrow}(p_-,\omega_-)}{\mathcal{D}_{AR}\mathcal{D}_{RR}}+\\
+\frac{G_{A}^{
\uparrow}(p_-,\omega_-)G_{K}^{\downarrow}(p_+,\omega_+)}{\mathcal{D}_{AR}\mathcal{D}_{AA}}]
\end{eqnarray}
where subscripts $p_{\pm} = (p\pm q/2)^2/2m$ and $\omega_{\pm} =
\omega\pm\Omega/2$ define the electrons momentum and frequency,
while functions $\mathcal{D}_{XY}$ are given as
\begin{equation}
\mathcal{D}_{XY} = 1-
\frac{1}{2\pi\nu\tau}\int\frac{dp}{(2\pi)^3}G_{X}^{\uparrow}(p_-,\omega_-)
G_{Y}^{\downarrow}(p_+,\omega_+)
\end{equation}
Evaluating expression (\ref{suscept}), we suggest the exchange
coupling $\lambda(p)$ to be constant. Further we will discuss the
dependence of $\lambda(p)$ on momentum and obtain the
corresponding spin wave's dispersion.

The spin polarization $\mathcal{P}$ in the paramagnetic region is
defined as
\begin{equation}\label{R-def}
\mathcal{P} = -\frac{i}{2N}\int \frac{d\omega
d\mathbf{p}}{(2\pi)^4}[G_{K}^{\uparrow}(\mathbf{p},\omega)
-G_{K}^{\downarrow}(\mathbf{p},\omega)]
\end{equation}
where $N$ is the total electron concentration. Equation
(\ref{R-def}) is equivalent to condition
$1-\lambda\nu\Pi(0,\Omega=2\omega_z)=0$. In equilibrium case at
$\omega_z=0$ it determines paramagnet-ferromagnet transition
point.

In general, polarization $\mathcal{P}$ consists of equilibrium
part induced by Zeeman effect and non-equilibrium part appearing
due to $\delta\mu$ shift. The non-equilibrium part of polarization
can be either positive or negative depending on the direction of
the injection current. It is convenient to introduce the energy
corresponding to $\mathcal{P}$
\begin{equation}\label{Om-def}
\Omega_{ex}=\lambda N\mathcal{P}
\end{equation}
The value $|\Omega_{ex}|<<\mu$ is the exchange energy of
electrons. Given the Zeeman splitting $2\omega_z$ and the shift
$\delta\mu$ of quasi-chemical potential, one obtains from
(\ref{GK}) and (\ref{R-def}) the equation for polarization, or
equivalently for $\Omega_{ex}$
\begin{equation}\label{condition}
\Omega_{ex}=\lambda\nu(\Omega_{ex}+ \delta\mu +
2\omega_z)\left[1-\frac{1}{6}\left(\frac{\Omega_{ex}+ \delta\mu +
2\omega_z}{2\mu}\right)^2\right]
\end{equation}
Note that we used the particle conservation condition under the
spin injection and the parabolic electron dispersion in deriving
the Eq. (\ref{condition}).

The spontaneous polarization vanishes at $\lambda\nu < 1$ and Eq.
(\ref{condition}) shows that the sign of $\Omega_{ex}$ coincides
with the sign of $\delta\mu + 2\omega_z$ and in the $\lambda\nu\ll
1$ limit one obtains
\begin{equation}\label{weak}
\Omega_{ex}=\frac{\lambda\nu}{1-\lambda\nu}(\delta\mu +
2\omega_{z})
\end{equation}
At $1-\lambda\nu \ll 1$ the exchange energy is much larger than
$|\delta\mu + 2\omega_{z}|$ and from Eq. (\ref{condition}) it
follows that
\begin{equation}\label{condition0}
\delta\mu + 2\omega_z \approx \Omega_{ex}\frac{1}{6}
\left(\frac{\Omega_{ex}}{2\mu}\right)^2
\end{equation}
If the Stoner criterion of ferromagnetism $\lambda\nu > 1$ is
realized then Eq. (\ref{condition}) has the nonzero solution even
in the absence of the external field and pumping \cite{bib:Moriya}
\begin{equation}
\frac{1}{6}\left(\frac{\Omega_{ex}}{2\mu}\right)^2=1-\frac{1}{\lambda\nu}
\end{equation}
\begin{figure}[t] \centering
\includegraphics[width=8cm]{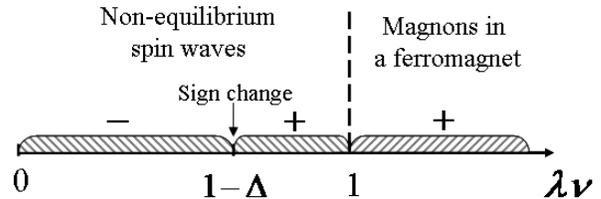}
\caption{$\lambda\nu=1$ represents the ferromagnet - paramagnet
phase transition point on line $0<\lambda\nu<\infty$. Point
$\lambda\nu =1-\Delta$ where $\Delta
=\frac{1}{3}(|\delta\mu|/4\mu)^{2/3}$ corresponds to the sign
change of the non- equilibrium spin waves spectrum in the limit
$|\Omega_{ex}|\tau>1$}\label{fig:2}
\end{figure}

\section{Gapless spin waves} Now let us discuss the spin waves
spectrum dependence in the regimes of both spontaneous and
non-equilibrium polarizations. The spectrum of spin wave
excitations is defined by the poles of the dynamical magnetic
susceptibility $\chi^{+-}(q,\Omega)$ in Eq. (\ref{suscept}),
namely $ 1 - \lambda \nu\Pi(q,\Omega)=0. $

We obtain the expression for the real part of spectrum in the
limit $|\Omega_{ex}|\tau>1$ at small momentum and frequency
$|\Omega_{ex}| \gg qv_F, |\Omega|$, where $v_F$ is the Fermi
velocity, the expression
\begin{eqnarray}\nonumber \label{spectr_pure}
\Re\Omega=2\omega_z\textrm{sign}(\mathbf{n}\mathbf{H}) -
\frac{\lambda\nu}{3}\left(\frac{qv_F}{\Omega_{ex}}\right)^2[\delta\mu + 2\omega_z\\
- (\Omega_{ex} - (\omega_z+\delta\mu/2))\left(\frac{\Omega_{ex}
+\delta\mu + 2\omega_z}{4\mu}\right)^2]
\end{eqnarray}
where $\mathbf{n}$ is the unit vector in the direction of the
magnetization in the paramagnetic metal. This expression should be
supplemented with Eq. (\ref{condition}) in order to obtain the
spectrum for the different regimes of pumping and values of the
exchange interaction $\lambda\nu$.

The Zeeman term in Eq. \ref{spectr_pure} has negative sign so long
as the direction of magnetization in the paramagnetic metal
$\mathbf{n}$ is opposite to the direction of the magnetic field
$\mathbf{H}$. The spectrum of spin waves becomes gapless in the
limit $\omega_z=0$. Under this circumstance the polarization in
the paramagnet is defined by the spin injection, $\delta\mu$,
only.

\section{Spin waves frequency sign change} In this section, we
consider the case where $\omega_z =0$. Firstly, assume that the
exchange coupling $\lambda \nu > 1$ and the system is in the
ferromagnetic state. Using Eq. (\ref{condition}) one sees that
$\Omega_{ex}\gg \delta\mu$ and the nonlinear term in the square
brackets of Eq. (\ref{spectr_pure}) is predominant. Taking into
account the dissipation one obtains the magnon spectrum in the
ferromagnet \cite{bib:Moriya}
\begin{equation}\label{spectr_ferr}
\Omega =
\frac{\lambda\nu}{3}\left(\frac{qv_F}{4\mu}\right)^2\left(|\Omega_{ex}|
- \frac{\lambda\nu Dq^2}{|\Omega_{ex}|\tau -i} \right)
\end{equation}
where $D= v^2_F \tau/3$ is the diffusion coefficient. The real
part of the spectrum is given by Goldstone mode, while the damping
of spin wave is very small and proportional to the quartic term of
the wave vector \cite{bib:Goldstone}.

Now let us tern to the paramagnetic side, $\lambda\nu <1$, of the
ferromagnet - paramagnet phase transition when the spin
polarization is induced due to spin injection. In the vicinity of
phase transition point
\begin{equation}
1-\lambda \nu \ll 1
\end{equation}
the sign of the wave-vector dependent part of the spectrum, Eq.
(\ref{spectr_pure}), stays positive as it is in the ferromagnetic
phase. However, we find the sign change crossover of q-dependent
part of the spectrum below some critical value of $\lambda\nu$
(see also Fig. \ref{fig:2}). Indeed, in this regime Eqs.
(\ref{condition}, \ref{condition0}) reduce to
\begin{equation}\label{condition1}
\delta\mu \approx \Omega_{ex}\frac{1}{6}
\left(\frac{\Omega_{ex}}{4\mu}\right)^2
\end{equation}
Combining Eq. (\ref{condition1}) with Eq. (\ref{spectr_pure}) one
sees that when parameter $\lambda\nu$ is smaller than the value
\begin{equation}
\lambda\nu \approx
1-\frac{1}{3}\left(\frac{|\delta\mu|}{4\mu}\right)^{2/3}
\end{equation}
the wave-vector dependent part of the spectrum in Eq.
(\ref{spectr_pure}) changes sign and becomes negative. More
generally, the width of the positive frequencies region on the
paramagnetic side of the phase transition strongly depends on the
details of the free electron dispersion, which was assumed
parabolic.

\begin{figure}[t] \centering
\includegraphics[width=8cm]{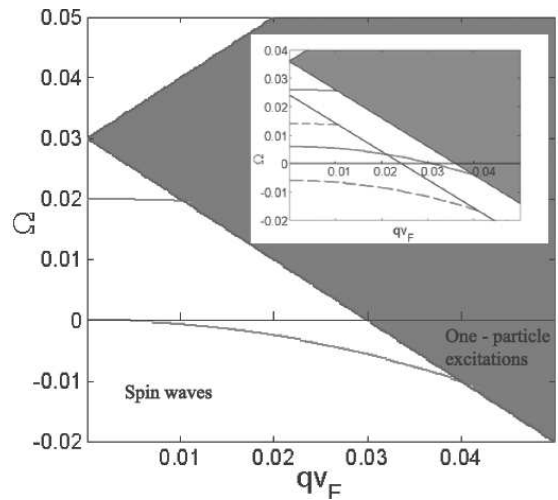}
\caption{Non-equilibrium spin waves dispersions of soft and gapped
modes for $\omega_z=0$ in $|\Omega_{ex}|\tau>1$ limit. Here all
values are measured in units of $\mu$: $|\Omega_{ex}|=0.03$,
$\lambda\nu=0.1$, $|\Omega_{ex}|\tau=1.3$. Filled area
$-qv_F>\Omega - |\Omega_{ex}|>qv_F$ corresponds to one-particle
excitations. Inset: Gapped modes with account of Zeeman effect in
$\omega_z=0.003$ in $|\Omega_{ex}|\tau>1$ limit. Solid lines
correspond to $\mathrm{sign}(\mathbf{nH})>0$ and dashed lines to
$\mathrm{sign}(\mathbf{nH})<0$ regimes. }\label{fig:3}
\end{figure}

Finally, in the limit of small $\lambda \nu \ll 1$ when the
exchange energy of electrons is small compared to $\delta\mu$ and
$2\omega_z$, we recover the result of Aronov \cite{bib:Aronov}
\begin{equation}\label{spectrAronov}
\Omega \simeq -\frac{(qv_F)^2}{3|\Omega_{ex}|}
\end{equation}
Here $|\Omega_{ex}|$ is defined by Eq. (\ref{weak}) at
$\omega_z=0$. This dependence is presented in Fig. (\ref{fig:3}).
Note that the frequency is inversely proportional to the exchange
coupling and has the opposite sign compared to the case of
equilibrium ferromagnet, Eq. (\ref{spectr_ferr}).

Accounting for electron scattering by impurities and Zeeman energy
gives expression in limit $\lambda \nu \ll 1$
\begin{equation}\label{spectr_para}
\Omega  = 2\omega_z \textrm{sign}(\mathbf{n}\mathbf{H}) -
\frac{(1-\lambda \nu)Dq^2}{|\Omega_{ex}|\tau - i}
\end{equation}
Here the value of the exchange energy $|\Omega_{ex}|$ is given by
Eq. (\ref{weak}). Here we do not impose limitations on
$|\Omega_{ex}|\tau$. The dispersion Eq. (\ref{spectr_para}) in the
limit $\Omega_{ex}\tau>1$ is shown in Fig. (\ref{fig:3}). Inset
illustrates regimes when the magnetization in the paramagnetic
metal is in the same $\mathrm{sign}(\mathbf{nH})>0$ or in the
opposite direction $\mathrm{sign}(\mathbf{nH})<0$ to the magnetic
field. The damping of spin-wave excitations is defined by the
imaginary part of the frequency $\Omega$, which in the limit
$|\Omega_{ex}|\tau\gg1$ is much smaller than the real part of
$\Omega$. The damping of the spin-wave excitations increases in
the limit $|\Omega_{ex}|\tau<1$ and is defined by the diffusion
coefficient modified by exchange interactions by a factor of
$1-\lambda \nu$.

In the case of the equilibrium regime in the absence of the spin
injection, $\delta \mu=0$, Eq. (\ref{spectr_para}) coincides with
the results of \cite{bib:Fulde, bib:Plaz}. Note that in the
equilibrium regime first Zeeman term in Eq. (\ref{spectr_para}) is
positive and as a result the real part of the total frequency is
also positive. Actually, the gapped Zeeman mode is related to the
total spin precession around the external magnetic field
$\mathbf{H}$. Note that the direction of the total magnetization
in the paramagnet depends both on the non-equilibrium
magnetization produced by spin injection and on the equilibrium
part due to external magnetic field. Naturally, the total
magnetization can be either in the direction of the applied
magnetic field or in the opposite direction.

Let us estimate the typical values of the wave vector $q$ and the
frequency $|\Omega|$ of the spin waves. In order to investigate
the properties of the non-equilibrium spin waves, the typical
length of the magnetic system $L=1/q$ has to be smaller than the
value of the spin-diffusion length $\ell_s$. Otherwise, the
propagation of the spin waves in the paramagnet will be strongly
suppressed by the spin-flip scattering.

Due to electrical current limitation it is difficult to achieve
high enough non-equilibrium spin polarization in metallic system.
The possible candidate to observe the non-equilibrium spin waves
might be a semiconductor structure, where polarization of order
$\Omega_{ex}/\mu\sim 1$ at $\Omega_{ex}$ in $\mathrm{meV}$ range
can be created (\cite{bib:Semi1} and references therein). At Fermi
velocity  $v\sim 10^6 \mathrm{cm/s}$ maximum $q<10^6
\mathrm{cm^{-1}}$ might be much larger than inverse spin-diffusion
length $\ell_s \sim1 \mathrm{\mu m}$. Which is reasonable value
for semiconductor structure. At $q \sim 10^4 \mathrm{cm^{-1}}$ we
estimate $\Omega$ in  $\mathrm{GHz}$ region.

\section{Gapped modes} Now let us discuss the dependence of
exchange coupling $\lambda(\mathbf{p}-\mathbf{k})$ on the
momentum. Setting $p$ and $k$ to Fermi momentum $p_F$ one can
expand the interaction function over Legendre polynomials
\cite{bib:White} as
\begin{equation}\label{expand}
\lambda(\mathbf{p}-\mathbf{k}) =
\sum_{\ell\geq0}\lambda_{\ell}P_{\ell}(\cos\theta)
\end{equation}
where $\theta$ is the angle between the vectors $\mathbf{p}$ and
$\mathbf{k}$. In the zero wave vector case, $q=0$, the transverse
magnetic susceptibility diverges if the following condition is
satisfied
\begin{equation}
1-\lambda_{\ell}\Pi_{\ell}(q=0,\Omega_{\ell}) = 0
\end{equation}
As a result, the spectrum of the spin waves is given by the series
of gapped modes both in ferromagnet and paramagnet regimes
\begin{equation}
\Omega_{\ell}(q=0) =
2\omega_z\textrm{sign}(\mathbf{n}\mathbf{H})+|\Omega_{ex}|(1-\lambda_{\ell}/\lambda)
\end{equation}
where $\ell=1,2...$ and $\lambda$ is related to the $\ell=0$ case.
The gapped modes are proportional to the polarization modified by
the factor, depended on the interaction constants. Naturally, in
the absence of the applied magnetic field, at $\mathrm{H}=0$,
these gapped modes are related to the processes of the spin
precession around the internal effective magnetic field originated
from the electron exchange interactions.

For simplicity, we will calculate the wave-vector dependent part
of the first mode ($\ell=1$) only. The corresponding dispersion
equation of the gapped non-equilibrium spin waves is given as
\begin{equation}
\Omega_{1} =2\omega_z\textrm{sign}(\mathbf{n}\mathbf{H})+
|\Omega_{ex}|(1-\lambda_{1}/\lambda)-
\frac{9\lambda}{5\lambda_{1}}\frac{(1-\lambda_{1}\nu)Dq^2}
{|\Omega_{ex}|\tau-i\lambda/\lambda_{1}}
\end{equation}
and presented in Fig. (\ref{fig:3}). The propagation of the gapped
spin waves is also suppressed in the $|\Omega_{ex}|\tau<1$ limit,
while the relaxation time is reduced by the value of
$\lambda_1/\lambda$ compared to the gapless spin wave in the
paramagnetic regime, Eq. (\ref{spectr_para}). We also note, that
$q$-dependent part of the spectrum does not change sign at
ferromagnet-paramagnet transition contrary to the gapless mode.

\section{Amplification of the electromagnetic field} Consider the interaction of the
electromagnetic field with the non-equilibrium spin-wave system.
The energy of the field will be supplied to the system when the
spin waves with positive frequency are excited. On the other hand,
the energy is extracted from the system in the case of negative
spin wave frequency. In this regime the electromagnetic field
interacting with the system might be amplified. Indeed, this
statement can be verified by estimating the energy dissipation of
the electromagnetic field in the spin system.

Experimentally, one might examine the absorbtion in the spin-wave
resonance effect \cite{bib:White, bib:Spinwave}, in the vicinity
of the spin waves resonant frequency $\Omega$, where wave vector
$q$ is defined by the geometry of the sample. One obtains for the
energy dissipation of the field
\begin{eqnarray}\nonumber\label{dissipation}
&Q&=\frac{1}{4\pi}\left<H(t)\frac{\partial B(t)}{\partial t}\right> =\\
&=&|H|^2\frac{\omega }{8} \Im[
\chi^{+-}(q,\omega)+\chi^{-+}(q,\omega)]
\end{eqnarray}
The transverse susceptibility $\chi^{+-}(q,\omega)$ is given by
Eq. (\ref{suscept}) and can be written as
\begin{equation}\label{suscept_res}
\chi^{+-}(q,\omega)=-2\mu_{B}^{2}\frac{N \mathcal{P}}{\omega -
\Re\Omega + i\gamma}
\end{equation}
where $N$ is the electron concentration, $\mathcal{P}>0$ is the
spin polarization, while $\Re\Omega$ is the real part of the
spin-wave frequency given by Eq. (\ref{spectr_pure}), $\gamma>0$
is the damping of spin waves in the paramagnetic metal, see Eq.
(\ref{spectr_para}), and in the ferromagnetic metal, Eq.
(\ref{spectr_ferr}). The susceptibility $\chi^{-+}(q,\omega)$
differs from Eq. (\ref{suscept_res}) by the opposite sign at
$\omega + i\gamma$. The imaginary part in Eq. (\ref{dissipation})
is
\begin{eqnarray}\nonumber
\Im\left[\chi^{+-}(q,\omega) + \chi^{-+}(q,\omega)\right]=\\
=2\mu_{B}^{2}\left[\frac{N \mathcal{P}\gamma}{(\omega -
\Re\Omega)^2 + \gamma^2}-\frac{N \mathcal{P}\gamma}{(\omega +
\Re\Omega)^2 + \gamma^2} \right]
\end{eqnarray}
Finally, in the vicinity of the resonant frequency $\Omega$ we
estimate the resulting expression for the energy dissipation
\begin{equation}
Q\sim \mu^2_B|H|^2 \frac{N \mathcal{P} \Re\Omega}{\hbar\gamma}
\end{equation}
The real part of the frequency $\Re\Omega$ in the ferromagnet is
always positive, Eq. (\ref{spectr_ferr}). Thus, the
electromagnetic field contributes the energy to the spin system,
which reveals in the well known spin-wave resonance effect, where
one observes the microwave absorbtion peaks by the standing spin
waves in thin ferromagnetic films. Note that the spin-waves'
frequency is also positive in the vicinity of the
paramagnet-ferromagnet phase transition $|1-\lambda \nu|\ll 1$ of
the non-equilibrium paramagnet, Fig. \ref{fig:2}. In this regime
the value of the dissipation, $Q>0$, and the energy of the
electromagnetic field is still supplied to the spin-wave system.

However, the energy influx to the magnetic system becomes negative
$Q<0$ when the frequency changes sign $\Re\Omega<0$, Eq.
(\ref{spectr_para}), at $\lambda\nu < 1$. Therefore, the electron
exchange interactions in the paramagnetic material with the
non-equilibrium spin polarization can lead to the amplification of
electromagnetic field. Experimentally, one might suggest the
combination of the spin-wave resonance effect with the spin
injection method. We propose the observation of the energy dips in
the absorption spectrum instead of the expected absorption peaks
in the spin-resonance effect.

\section{Conclusions}
To conclude, we have studied the effect of electron exchange
interactions on the non-equilibrium spin excitations in the
paramagnetic metal. We have obtained the energy spectrum of the
non-equilibrium spin waves in the diffusive paramagnetic system.
Generally speaking, the negative imaginary part of the spin wave
frequency shows the stability of the non-equilibrium state under
small spin-wave perturbations. In the regime when
electron-electron exchange energy is larger than the inverse
electron mean free time the resulting spin waves are long-lived
excitations, while in the diffusive limit the damping of spin
waves is very strong. The excitation of these spin waves lowers
the energy of the system due to negative sign of the real part of
the frequency. Interestingly, we have shown that the
electromagnetic field can be amplified in the non-equilibrium spin
polarized paramagnetic metal. The spin injection plays a role of
pumping source establishing the long-lived collective spin
excitations which form an active medium leading to the
amplification of the electromagnetic field.

We are grateful for the financial support of RFFI under Grant
No.10-02-00681-A and Federal Programm under Grant No.
2009-1.5-508-008-012.

\end{document}